\documentclass[10pt]{article}

\usepackage{graphicx}
\usepackage{dcolumn}
\usepackage{bm}

\begin{document}
\begin{center}
{{\LARGE \bf QCD Phase Transition in a new Hybrid Model Formulation }} \\
\bigskip
{\large \bf P. K. Srivastava$^*$\footnote{email: prasu111@gmail.com}} {\large \bf C. P. Singh$^*$} \\
{$^*$Department of Physics, Banaras Hindu University, Varanasi 221005, India} 
\bigskip
\end{center}

\begin{abstract}
Search of a proper and realistic equations of state (EOS) for strongly interacting matter used in the study of QCD phase diagram still appears as a challenging task. Recently, we have constructed a hybrid model description for the quark gluon plasma (QGP) as well as hadron gas (HG) phases where we use a new excluded-volume model for HG and a thermodynamically-consistent quasiparticle model for the QGP phase. We attempt to use them to get a QCD phase boundary and a critical point. We test our hybrid model by reproducing the entire lattice QCD data for strongly interacting matter at zero baryon chemical potential ($\mu_{B}$)and predict the results at finite $\mu_{B}$ and $T$. 
\end{abstract}


\section{Introduction}
Quantum chromodynamics (QCD) predicts that at sufficiently high temperatures ($T$) and/or chemical potentials ($\mu_{B}$), strongly interacting matter goes through a phase transition from colour insulating hadron gas (HG) phase to colour conducting quark gluon plasma (QGP) phase ~\cite{a,b}. Heavy ion collisions can provide a unique opportunity to study this QCD phase transition from HG to QGP. Our endeavour in such studies is the search of a suitable equation of state (EOS) for the description of both phases of strongly interacting matter. Significant success has been gained in lattice calculations using QCD thermodynamics to provide a valid EOS for QCD matter at zero baryon chemical potential but at large $T$. However, the lattice methods are unreliable to describe the properties of matter at finite density of baryons. Therefore, finding an EOS for QCD matter valid at zero as well as non-zero chemical potential is a challenging problem. In this paper, we construct a  hybrid model where a new excluded-volume model is proposed for HG description and a thermodynamically-consistent quasiparticle model is given for the QGP phase in predicting the properties of entire QCD matter. Moreover, we use the hybrid model to get the QCD phase boundary, location of critical point (CP) and determine the order of the phase transition. 
\section{EOS for QGP and HG}
The EOS for QGP in our quasiparticle framework and the calculations regarding thermodynamical quantities like pressure, energy density, particle density etc. can be found in our earlier work ~\cite{c,d}. In this model, we start with the definition of average energy density and average number density of particles and derive all other thermodynamical quantities from them in a consistent manner. The parameters used in the calculations are given there. 

 Recently we proposed a new excluded-volume model for the hot and dense HG ~\cite{c,e,f}. The grand canonical partition function in our excluded volume model of HG can be explicitly given as follows :

\begin{equation}
ln Z_i^{ex} = \frac{g_i}{6 \pi^2 T}\int_{V_i^0}^{V-\sum_{j} N_j V_j^0} dV\int_0^\infty \frac{k^4 dk}{\sqrt{k^2+m_i^2}} \frac1{[exp\left(\frac{E_i - \mu_i}{T}\right)+1]}
\end{equation}
where $g_i$ is the degeneracy factor of ith species of baryons,$E_{i}$ is the energy of the particle ($E_{i}=\sqrt{k^2+m_i^2}$), $V_j^0$ is the eigenvolume assigned to each baryon of ith species and hence $\sum_{j}N_jV_j^0$ becomes the total occupied volume where $N_{j}$ represent the total number of jth baryons. 

\section{Results and Comparison with Lattice QCD}

\begin{figure}
\begin{center}
\includegraphics[height=15em]{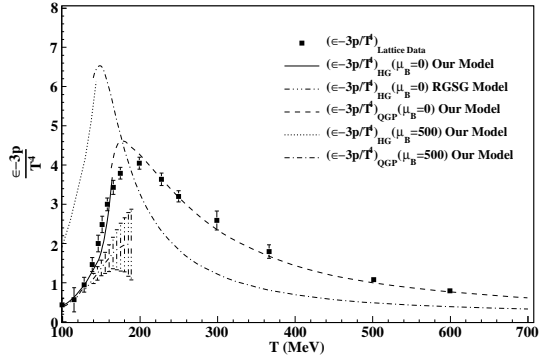}
\caption[]{Variations of trace anomaly $\left(\epsilon-3p\right)/T^{4}$ with respect to temperature at $\mu_{B}=0$ and $500$ MeV in our hybrid model. Shaded portion represents the calculation of Andronic et al.~\cite{g}.}
\end{center}
\end{figure}

\begin{figure}
\begin{center}
\includegraphics[height=15em]{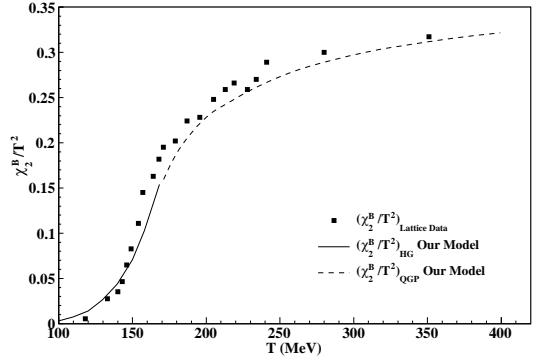}
\caption[]{Variation of normalized baryon number succeptibility with respect to temperature at $\mu_{B}=0$ in our hybrid model. Lattice data points at $\mu_{B}=0$ are taken from Ref.~\cite{h}.}
\end{center}
\end{figure}

\begin{figure}
\begin{center}
\includegraphics[height=15em]{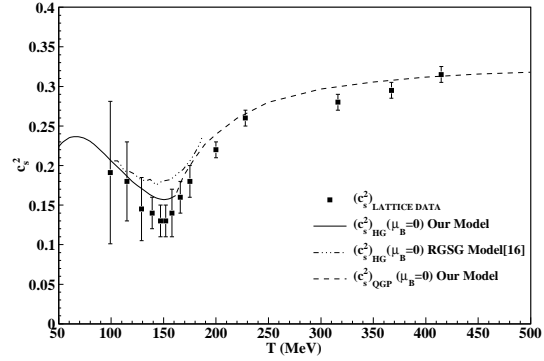}
\caption[]{Variation of speed of sound with the temperature at $\mu_{B}=0$ MeV as obtained in our hybrid model and compared with lattice results~\cite{i,j}. Dash-tripple dotted curve shows the result of Andronic et al.~\cite{g}.}
\end{center}
\end{figure}

In Fig. 1, we have plotted the results obtained for the trace anomaly factor $\left(\epsilon-3p\right)/T^{4}$ in our hybrid model calculations using HG and QGP equations of state separately at $\mu_{B}=0$. We further compare our results with the results obtained from a recent lattice calculation ~\cite{h}. We notice that our results yield an excellent fit to the lattice data.In Fig. 2, we show the variations of baryonic succeptibility normalized as $\chi^{B}_{2}/T^{2}$ with temperature at $\mu_{B}=0$ and compare our results with the lattice QCD results. Our hybrid model results again compare well with the lattice data points. Here again the curves, for $\chi^{B}_{2}/T^{2}$ obtained for HG and QGP phases are smoothly connected at around $T=170$ MeV, which shows the presence of a cross-over transition between two phases. In Fig. 3, we show the variations of square of speed of sound ($c_{s}^{2}$) with respect to temperature at $\mu_{B}=0$ MeV and again a comparison is given with the lattice QCD results. For $\mu_{B}=0$, our results reproduces the lattice results very well. We have also shown separately the curve for $c_{s}^{2}$ obtained by Andronic et. al. by using RGSG model for HG phase only ~\cite{g}. However, the features of the curve differ from the lattice data, although it also yields a minimum at around the same temperature.

\section{Critical Point (CP) and Order of Phase Transition}

 In this section we attempt to show the precise location and nature of CP existing on this phase boundary.

\begin{figure}
\includegraphics[height=20em]{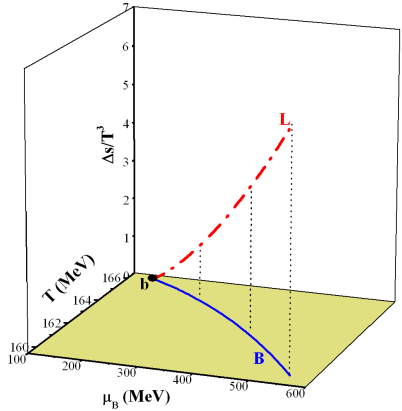}
\caption[]{Variation of $ (\Delta s/T^{3})=(s/T^{3})_{QGP}- (s/T^{3})_{HG}$ with respect to coordinates of various phase transition points on the $(T, \mu_{B})$ phase boundary. We have used transition points from Ref.~\cite{d}.}
\end{figure}

In Fig. 4, we show what will happen to a quantity depicting the change in the entropy density from the phase transition at CP of the phase diagram. We define the normalized difference $\frac{\Delta s}{T^{3}}=(s/T^{3})_{QGP}-(s/T^{3})_{HG}$ and demonstrate its variations with respect to the coordinates of the phase transition points lying at the deconfining phase boundary. We find that $\frac{\Delta s}{T^{3}}\ne 0.0$ and positive along the deconfining phase boundary in the case of first order phase transition which supports the role of the presence of a nonvanishing latent heat in the phase transition from HG to QGP. However, we surprisingly notice that $\frac{\Delta s}{T^{3}}\approx 0$ exactly at the CP and thus CP can be taken as a point where the first order phase boundary terminates and phase transition changes its order.

\section{Summary}
 Thus above results give a firm indication that the order of phase transition changes at CP for the deconfining phase transition. We hope that our results will clarify the mist surrounding the understanding of the deconfining phase transition and the conjectured phase boundary. More importantly, we have formulated a phenomenological hybrid model which provides a realistic EOS for the entire QCD matter and in the absence of a first-principle lattice QCD calculation especially at finite $\mu_{B}$, it can be reliably used for deriving the information on the QCD phase boundary.\cite{a,b}.\\

\section{Acknowledgement}
 PKS is grateful to the University Grants Commission (UGC), New Delhi for financial assistance.

\end{document}